\documentstyle[aps,preprint]{revtex}
\begin{document}
\title{Herd Formation and Information Transmission in a Population:
Non-universal behaviour}
\author{Dafang Zheng $^{1,2,}$\thanks
{e-mail: phdzheng@scut.edu.cn}, P. M. Hui $^{3}$, K. F. Yip $^{3}$
and N. F. Johnson$^{4}$}
\address{$^{1}$ Department of Mathematical Sciences, Brunel University,
Uxbridge, UB8 3PH, UK}
\address{$^{2}$ Department of Applied Physics, South China University of
Technology,\\ Guangzhou 510641, P.R. China}
\address{$^{3}$ Department of Physics, The Chinese University of Hong
Kong,\\
Shatin, New Territories, Hong Kong}
\address{$^{4}$ Department of Physics, University of Oxford,\\
Clarendon Laboratory, Oxford OX1 3PU, UK} \maketitle

\begin{abstract}
We present generalized dynamical models describing the sharing of
information, and the corresponding herd behavior, in a population
based on the recent model proposed by Egu\'{\i}luz and Zimmermann
(EZ) [Phys. Rev. Lett. {\bf 85}, 5659 (2000)]. The EZ model, which
is a dynamical version of the herd formation model of Cont and
Bouchaud (CB), gives a reasonable model for the formation of
clusters of agents and for actions taken by clusters of agents.
Both the EZ and CB models give a cluster size distribution
characterized by a power law with an exponent $-5/2$.  By
introducing a size-dependent probability for dissociation of a
cluster of agents, we show that the exponent characterizing the
cluster size distribution becomes model-dependent and
non-universal, with an exponential cutoff for large cluster sizes.
The actions taken by the clusters of agents generate the price
returns, the distribution of which is also characterized by a
model-dependent exponent.  When a size-dependent transaction rate
is introduced instead of a size-dependent dissociation rate, it is
found that the distribution of price returns is characterized by a
model-dependent exponent while the exponent for the cluster-size
distribution remains unchanged. The resulting systems provide
simplified models of a financial market and yield power law
behaviour with an easily tunable exponent.

\vspace*{0.5 true in}

\noindent PACS Nos.: 05.65.+b, 87.23.Ge, 02.50.Le, 05.45.Tp

\end{abstract}

\newpage
\section{Introduction}
Herd behaviour and information transmission among agents are
obviously important in financial markets. With the newly
established area of econophysics\cite{stanley,bouchaud}, various
multi-agent models of financial markets have been
studied\cite{farmer,lux}. In agent-based models, markets are
typically modelled by a population of competing agents.  Through
information transmission and opinion sharing, these agents may not
act independently. The collective behaviour of clusters of agents,
also referred to as crowds, in which there is efficient
information sharing among the agents is an important factor in
both real and simulated markets\cite{us1}. These crowds are
dynamic in nature in that there is a continual process of crowd
formation and dissociation.

Recently, Egu\'{\i}luz and Zimmermann\cite{EZ} proposed and
studied a simple model for stochastic opinion cluster formation
and information dispersal.  It is a dynamical model (henceforth
referred to as the EZ model) in which there is a continual
grouping and re-grouping of agents to form clusters. A cluster or
crowd of agents act together (either buying or selling) and then
dissolve after the transaction has occurred. When a cluster of
agents decides not to trade, i.e., an inactive state, it may
combine with another cluster of agents to form a bigger cluster.
Detailed numerical studies\cite{EZ} and mean field
analysis\cite{dHR1} revealed several interesting features of the
model.  For example, it was observed that the model could lead to
a fat-tail distribution of price returns qualitatively similar to
that observed in real markets\cite{stanley,gopikrishnan}. In
addition, the cluster size distribution $n_{s}$ shows a scaling
behaviour of the form $n_{s} \sim s^{-5/2}$ for a range of cluster
size $s$, followed by an exponential cutoff\cite{dHR1}. The EZ
model represents a dynamical version of the static
percolation-type model of Cont and Bouchaud\cite{cont} in which
herd formation is described by random connection between agents,
and the cluster size distribution is found to follow the same
scaling behaviour and with the same value of $-5/2$ for the
exponent.  Several variations on the model have been proposed and
studied.  These variations include the spreading of opinion to
multiple clusters \cite{dHR1} and inhomogeneous dissociation of
clusters\cite{dHR2}. Interestingly, it was found that the values
of the exponent characterizing the cluster size distribution seem
to be robust against these variations, i.e., these values remain
unchanged for the different variations proposed so far.  We note
that similar scaling behavior has been found in the size
distribution of businesses\cite{Ramsden,Nagel}, although the value
of the exponent is different.

Here we introduce two generalized versions of the EZ model. One
version has a size-dependent probability for the dissociation of
clusters after opinion sharing.  The distributions of cluster
sizes and price returns are found to be characterized by
model-dependent exponents. Another version has a size-dependent
transaction rate.  It leads to a price returns distribution with a
model-dependent exponent, while keeping the exponent for the
cluster-size distribution to be $-5/2$.  Both models can be
treated analytically via a mean field approach.  The mean field
results are found to be in good agreement with numerical results.
The EZ model is re-covered as a special case of our generalized
models.

The plan of the paper is as follows.  We define our generalized
model with size-dependent cluster dissociation probability in
Sec. II and present the numerical results of the model.
Section III provides a mean field analysis.
The analytic results are compared with numerical simulations.
In Sec. IV, we define another generalized model with a size-dependent
transaction rate and present numerical results.  The results are
summarized in Sec. V.

\section{Model I: Size Department Cluster Dissociation Rate}

We consider a model with a total of $N$ agents.  Following Ref.\cite{EZ},
a cluster or crowd is a group of agents who can exchange
information efficiently.  These agents make
the same decision at a given moment in time.  Initially, all the agents
are isolated.  As
time evolves, an agent belongs to a cluster of a certain size.  At each
timestep an agent, say the $i$-th one, is chosen at random.  Let
$s_{i}$ be the size of the cluster to which the agent belongs.
Since the agents
within a cluster have a common opinion, all agents in such a cluster tend
to imitate each
other and hence act together.  With  probability $a$ the agent, and hence
the whole
cluster, decides to  make a transaction, e.g. to buy or to sell with equal
probability.
After the transaction, the
cluster is then broken up into isolated agents with
a probability $s_{i}^{-\delta}$, with $0 \leq \delta < 1$.  With
probability $(1-a)$ the agents decide not to make a transaction,
i.e., they wait and try to
gather more information. The other agents in the cluster follow.
In this case, another agent $j$ is chosen at random.  The two clusters
of sizes $s_{i}$ and $s_{j}$ then either combine to form a bigger cluster
with probability $s_{i}^{-\delta} s_{j}^{-\delta}$, or the two
clusters remain separate with probability $(1- s_{i}^{-\delta}
s_{j}^{-\delta})$.  Here $a$ can be treated as a parameter reflecting
the investment rate showing how frequent a transaction is made.  Our
model thus represents a generalization of the basic EZ model to the case
in which a cluster of agents may stay together to form a group
{\em after} making a transaction.
The probability of dissociation $s_{i}^{-\delta}$ implies
that larger clusters
have a larger tendency to remain grouped while smaller
clusters are easier
to break up\cite{dHR3}.  For the special case of $\delta = 0$, our model
reduces to the EZ model.

In the EZ model ($\delta = 0$), clusters of agents break up  after a
transaction.  Here, our model includes a dissociation  probability depending
on the
cluster size - this feature may be invoked to mimic practical aspects of a
financial
market, such as the effect of news arrival. Imagine one of
the agents in a cluster of size $s_{i}$ receives some external news
with probability $a$ at a given timestep.  This external news suggests
that he, and hence the other members of his cluster, should immediately
trade
(buy or sell).  Since the news is external, the crowd act together in this
one moment,
leaving the cluster with a finite probability of subsequently dissociating.
Suppose that they sense, e.g. from the resulting price-movement, that they
are a member of a large crowd of like-minded agents: in practice many
traders like to
feel part of a larger crowd for reassurance. We therefore assume that the
crowd breaks up
with a size-dependent probability $p(s_{i})$, where $p(s_{i}) = 1$ for
$s_{i}=1$ and
$p(s_{i})$  decreases monotonically as $s_{i}$ increases.  By contrast, with
probability
($1-a$) there  is no news arrival from outside.  The agent in the chosen
cluster,
uncertain about whether to buy or sell, makes contact with an agent in
another cluster of
size
$s_{j}$.  The agents share information and come up with a new opinion.  Each
of them then
separately tries to persuade the other members of his cluster of the new
opinion.  With
probability
$p(s_{i})$ ($p(s_{j})$)  the opinion of cluster $i$ ($j$) changes to the new
opinion.
Thus,  the two clusters combine with probability $p(s_{i})p(s_{j})$.  It
turns
out that this particular form of the two combined modifications to the EZ
model, can be treated analytically
using our mean field analysis.  As a specific example,
our
numerical  simulations are carried out for the case in which $p(s) \sim
s^{-\delta}$.

Let $n_{s}$ be the number of clusters of size $s$.  Figure 1 shows the
results of
numerical  simulations on the cluster size distribution in the steady state,
for various values of the parameter $\delta$.  The results are obtained for
a system with
$N=10^{4}$ agents and
$a=0.3$ \cite{EZ}.  Averages are taken over a time window of $10^{6}$
time steps
after the transient behavior has disappeared, together with a configuration
average over 100 different runs with different initial conditions.
The $\delta = 0$ results
give the features in the EZ model.  For a range of $s$, $n_{s} \sim
s^{-\beta}$ with $\beta = 5/2$ \cite{EZ}.  Deviation from the scaling
behavior sets in at a value of $s$ depending on the value of the
parameter $a$.  For smaller values of $a$,
the scaling region enlarges.
These features are consistent with previous
numerical\cite{EZ}
and analytical studies\cite{dHR1}.
For $0 \leq \delta < 1$, it is observed that
the size distribution $n_{s}$ still scales with $s$ in a range of $s$
as in the EZ model.  However, the exponent becomes model dependent
and hence {\em non-universal}.  The data shows that the exponent
$\delta$ is consistent with the behavior $n_{s} \sim s^{-\beta(\delta)}$,
where $\beta(\delta) = 5/2 - \delta$.  A mean field analysis can be
used to extract this scaling behavior, as will be described in the
next section.

It is interesting to note that several attempts have been made to
modify the EZ model.  These extensions include, for example, the
study by d'Hulst and Rodgers on democracy versus dictatorship by
incorporating an inhomogeneous investment rate in the
population\cite{dHR2} and also allowing rumor to spread to
multiple clusters in one time step after a chosen cluster decides
not to make a transaction\cite{dHR1}.  All the extensions proposed
so far give $\beta = 5/2$, hence the value seems to be robust. The
present model incorporates a size-dependent dissociation
probability of a cluster after a transaction and leads to a
tunable and model-dependent $\beta(\delta)$.  Thus our model
actually gives a set of models with different values of $\beta$,
similar to the case of changing a system from one universality
class to another in problems in critical phenomena.  In fact, the
situation is reminiscent of the non-universal exponent of
conductivity in continuum percolation\cite{Halperin,hui1}. In
percolation problems\cite{stauffer1}, it is known that the
effective conductivity for a system consisting of insulators and
conductors exhibits the scaling behavior $\sigma_{e} \sim
(p-p_{c})^{t}$ near the percolation threshold $p_{c}$.  The
exponent $t$ is universal in that its value depends only on the
dimension of the system, regardless of other details, e.g. lattice
type.  However, if the conductances $\sigma_{e}$ of the conductors
follow a distribution of the form $P(\sigma_{e}) \sim
\sigma_{e}^{-\delta}$ with $0 < \delta < 1$, the
$t$-exponent\cite{Halperin} and other related
properties\cite{hui1,hui2} become non-universal with exponents
taking on a value depending on $\delta$. It should be noted that
it is not so surprising to see a connection between percolation
and model for herd behavior. In the model of Cont and
Bouchaud\cite{cont}, the EZ model\cite{EZ} and their
variations\cite{dHR1,dHR2}, an agent could be connected to any one
of $(N-1)$ other agents to form a cluster.  These models hence
represent a problem of connectivity in high dimensions. Several
other percolation type models\cite{stauffer2,stauffer3,stauffer4}
have also been proposed to explain the phenomena observed in real
markets. It is also interesting to note that value of power law
exponents in statistical physics could also be made non-universal
by introducing long-range interactions.  By making the larger
clusters harder to dissociate, it can be thought of as effectively
introducing some long range interaction among agents in time.

Egu\'{\i}luz and Zimmermann\cite{EZ} applied their model to study
the distribution of price returns.  A price can be generated
according to
\begin{equation}
P(t+1) = P(t) \exp(s'/\lambda),
\end{equation}
where $\lambda$ is a parameter for the liquidity of the market.
The price return $R(t) = \ln P(t) - \ln P(t-1)$ is defined to be
the relative number of agents buying or selling at a time with $s'
= s$ for a cluster of agents deciding to buy, and $s' = -s$ for a
cluster deciding to sell at a given timestep.  Numerical results
for the EZ model showed that the distribution of returns $P(R)
\sim R^{-\alpha}$ with $\alpha = 3/2$.  We have carried out
similar calculations for our model.  Figure 2 shows the price
return distributions for different values of $\delta$ on a log-log
scales. As for the cluster size distribution, the exponent
$\alpha$ is now {\em non-universal} and takes on the value $3/2 -
\delta$, which is also the value of $\beta(\delta) -1$ \cite{EZ}.

\section{Mean field analysis}

The cluster size distribution in the EZ model can be studied via a
mean field analysis\cite{dHR1}.  The treatment can be extended to
the generalized model to extract the scaling behavior of $n_{s}$,
though the algebra is more complicated.  Let $n_{s}(t)$ be the
number of clusters of size $s$ at time $t$.  The dynamics of
$n_{s}(t)$ is governed by the following master equation describing
the result of collection action of the members of the cluster. The
equation is
\begin{equation}
N \frac{\partial n_{s}}{\partial t} =
-a s^{1-\delta} n_{s} + \frac{(1-a)}{N} \sum_{r=1}^{s-1}
r^{1-\delta} n_{r} (s-r)^{1-\delta} n_{s-r}
- \frac{2(1-a)s^{1-\delta} n_{s}}{N} \sum_{r=1}^{\infty} r^{1-\delta}n_{r}
\end{equation}
for $s > 1$.  Each of the terms on the right hand side of Eq.(2)
represents the consequence of a possible action of the agent.  The first
term describes the dissociation of a cluster of size $s$ after a
transaction is made.  The second term represents coagulation of two clusters
to form a cluster of size $s$.  The third term represents the
combination of a cluster of size $s$ with another cluster.  For clusters
of size unity ($s=1$), we have
\begin{equation}
N \frac{\partial n_{1}}{\partial t} =
a \sum_{r=2}^{\infty} r^{2-\delta}n_{r}
- \frac{2 (1-a)n_{1}}{N} \sum_{r=1}^{\infty} r^{1-\delta}n_{r}.
\end{equation}
Here, the first term comes from the dissociation of any clusters
into isolated agents and the second term describes the combination of a
cluster of size unity with another cluster.  In the steady state,
$\frac{\partial n_{s}}{\partial t} = 0$, we have
\begin{equation}
s^{1-\delta} n_{s} = A \sum_{r=1}^{s-1} r^{1-\delta}
(s-r)^{1-\delta} n_{r} n_{s-r}
\end{equation}
for $s>1$, and
\begin{equation}
n_{1} = B \sum_{r=2}^{\infty} r^{2-\delta} n_{r}
\end{equation}
for $s =1$, where
\begin{equation}
A = \frac{1-a}{Na + 2(1-a)\sum_{r=1}^{\infty} r^{1-\delta}n_{r}},
\end{equation}
and
\begin{equation}
B = \frac{Na}{2(1-a) \sum_{r=1}^{\infty} r^{1-\delta} n_{r}}.
\end{equation}

The aim here is to extract the scaling behaviour.  Invoking a generating
function approach\cite{wallace}, we let
\begin{equation}
G(\omega) = \sum_{r=0}^{\infty} r^{1-\delta} n_{r} e^{-\omega r} = g(\omega)
+
n_{1} e^{-\omega},
\end{equation}
where $g(\omega) = \sum_{r=2}^{\infty} r^{1-\delta} n_{r} e^{-\omega r}$.
It follows from Eq.(4) that the function $g(\omega)$ satisfies
the equation
\begin{equation}
g^{2}(\omega) + (2n_{1} e^{-\omega} - \frac{1}{A}) g(\omega)
+ n_{1}^{2} e^{-2\omega} = 0.
\end{equation}
Note that $A$ can be expressed in terms of $n_{1}$ and $g(0)$, and
\begin{equation}
g(\omega) = \frac{1}{4A} (1 - \sqrt{1 - 4n_{1} A e^{-\omega}})^{2}.
\end{equation}
The number of clusters of size $s$ can be found formally by
\begin{equation}
n_{s} = \frac{1}{s^{1-\delta} s!} \frac{\partial^{s}G}{\partial
z^{s}}|_{z=0},
\end{equation}
where $z = e^{-\omega}$.  The resulting expression for $n_{s}$ is
\begin{equation}
n_{s} = \frac{(2s-2)! (1-a)^{s-1} (\sum_{r=1}^{\infty}
r^{1-\delta}n_{r})^{s}
[(1-a) \sum_{r=1}^{\infty} r^{1-\delta} n_{r} + Na]^{s}}
{(s!)^{2} s^{-\delta} [Na + 2(1-a)\sum_{r=1}^{\infty} r^{1-\delta}n_{r}]
^{2s-1}}. \
\end{equation}
Invoking Sterling's formula yields
\begin{equation}
n_{s} \sim N \left[ \frac{ 4(1-a)[(1-a) + \frac{Na}{\sum_{r=1}^{\infty}
r^{1-\delta}n_{r}}]}
{[\frac{Na}{\sum_{r=1}^{\infty} r^{1-\delta}n_{r}} + 2(1-a)]^{2}}
\right]^{s}
s^{-(\frac{5}{2} - \delta)}.
\end{equation}
For $\delta = 0$, the sum $\sum_{r=1}^{\infty} r^{1-\delta} n_{r} = N$
and the previous results of Refs.\cite{EZ,dHR1} are recovered.
For $\delta \neq 0$,
it is difficult to solve for $n_{s}$.  Since the summations in Eqs.(12)
and (13) give a number, our result shows that
$n_{s} \sim s^{-\beta(\delta)}$ with $\beta(\delta) = 5/2 - \delta$
for the present model.  This $\delta$-dependent exponent is also
indicated in Fig. 1 (lines are a guide to the eye).
We note that the scaling behaviour is masked
by the behaviour of the term in the squared brackets in Eq.(13) for
large values of $s$, similar to the situation for the EZ model\cite{dHR1}.

\section{Model II: Size-dependent transaction rate}

In this generalization of the EZ model, we introduce a
size-dependent transaction rate instead of size-dependent
dissociation and combination rates of clusters. It is motivated by
the fact that it is hard to measure the structural properties of
herds, e.g. the size distribution.  Instead, the resulting
properties of the herd effect, e.g. price returns, are easier to
obtain\cite{stanley,gopikrishnan}.  Empirically, it was found that
the distribution of price returns, while taking on a power law
behaviour, has an exponent $\alpha$ that is deviated from the
value of $3/2$ in the EZ model.  It is therefore interesting to
investigate models in which the exponent for the distribution of
the price returns becomes tunable. Here, we introduce a model in
which the EZ dynamics is used for cluster formation and
dissociation, i.e., the cluster always dissolves after an action
is taken. However, when an agent $i$ is randomly chosen, there are
three possible actions. With probability $a(1 - s_{i}^{-\sigma})$,
the cluster of agents decide not to take an action (make a
transaction) and the cluster simply dissolves. With probability
$as^{-\sigma}$, the cluster of agents act collectively to either
buy or sell, and the cluster dissolves after the transaction is
made. With probability $(1-a)$, the chosen cluster combines with
another randomly picked cluster. For $\sigma = 0$, the present
model reduces to the EZ model.  For $\sigma \neq 0$, the model
reflects that while agents may have established connections to
exchange information through different means, they may tend to
look for common opinion before taking actions. Reaching a
consensus becomes difficult as the cluster size increases.
Therefore the effective transaction rate is lower than $a$ and
becomes size-dependent.

Since the combination and dissociation probabilities are identical
to those in the EZ model, the distribution of cluster sizes also
follows $n_{s} \sim s^{-5/2}$, with an exponential cutoff as in
the EZ model.  Figure 3 shows the distribution of price returns in
Model II for different values of $\sigma$ in a system with
$N=10^{4}$ agents and $a = 0.1$.  It is noted that the return
$P(R)$ follows a power law $P(R) \sim R^{-\alpha}$, where the
exponent $\alpha$ has become model-dependent.  The value of
$\alpha$ is well described by $\alpha = 3/2 + \sigma$.  The result
can be understood as the distribution of returns is given by the
product of the probability of selecting a cluster of size $s$
times the probability of making a transaction\cite{EZ}, i.e.,
$P(R) \sim R^{-\alpha} \sim s^{-5/2} s^{-\sigma} \sim s^{-(3/2 +
\sigma)}$. Since the price returns are generated by the collective
action of the whole cluster, we have $P(R) \sim R^{-(3/2
+\sigma)}$.

\section{summary}

In summary, we have proposed and studied the cluster size
distribution, and the price returns distribution, of two
generalized versions of the EZ model for herd behavior and
information sharing in a population.  We have carried out
numerical simulations on our models. The two generalizations can
be treated analytically within a mean field approach.  By
introducing a probability for dissociation of a cluster depending
on its size, the exponent characterizing the cluster size
distribution takes on a model-dependent non-universal value, with
a corresponding shift in the exponent characterizing the price
returns from the value of the EZ model.  It is also possible to
make the exponent for the distribution of price returns
model-dependent while keeping the exponent of the cluster-size
distribution unchanged by imposing a size-dependent transaction
rate. Our models thus illustrate that it is possible to tune the
power law behaviour.  It is interested to note that observed
values for the exponent $\alpha$ in markets can be as large as $4$
\cite{gopikrishnan}, which is quite different from the robust
value of $3/2$ in the EZ model. Within our model, a value of
$\alpha$ with $\alpha > 3/2$ can be attributed to the hesitation
among agents in a cluster in making a transaction.

Several extensions of our models are immediately possible.  Our
particular choices of the form of the probability for dissociation
of clusters in Model I and the probability for making a
transaction in Model II allow us to tune the exponent.  Exploring
other functional forms of the probabilities may alter the range of
values in which the exponent can be tuned. Our model can also be
extended to study the size distribution of businesses. A model
similar to the EZ model has already been proposed in this
context\cite{dHR3}. However the scaling behavior seems to be
non-universal for data from different countries\cite{Ramsden}. Our
work thus provides a possible extension to cope with this observed
non-universality.

\begin{center}
{\bf ACKNOWLEDGMENTS}
\end{center}

One of us (DFZ) acknowledges the
support from the China Scholarship Council and
the Natural Science Foundation of Guangdong Province, China.
Work at CUHK was supported in part by a Grant (CUHK4241/01P)
from the Research Grants Council of the Hong Kong SAR Government.

\newpage \centerline{\bf FIGURE CAPTIONS}

\bigskip
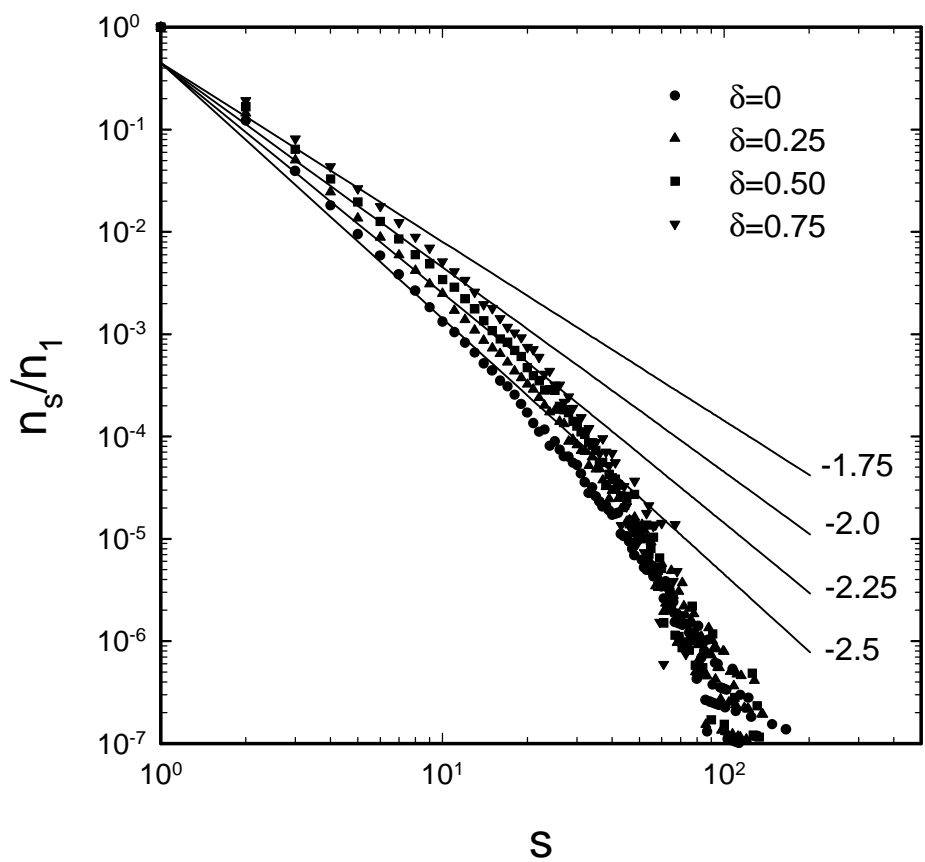
\noindent Figure 1: The size distribution $n_{s}/n_{1}$ as a function
of the size $s$ on a log-log scale for different values of
$\delta$ obtained by numerical simulations (symbols) of Model I.
The values of
$\delta$ used in the calculations are: $\delta = 0, 0.25, 0.50, 0.75$.
The solid lines are a guide to the eye corresponding to exponents
$\beta = 2.5, 2.25, 2.0, 1.75$ respectively.

\bigskip
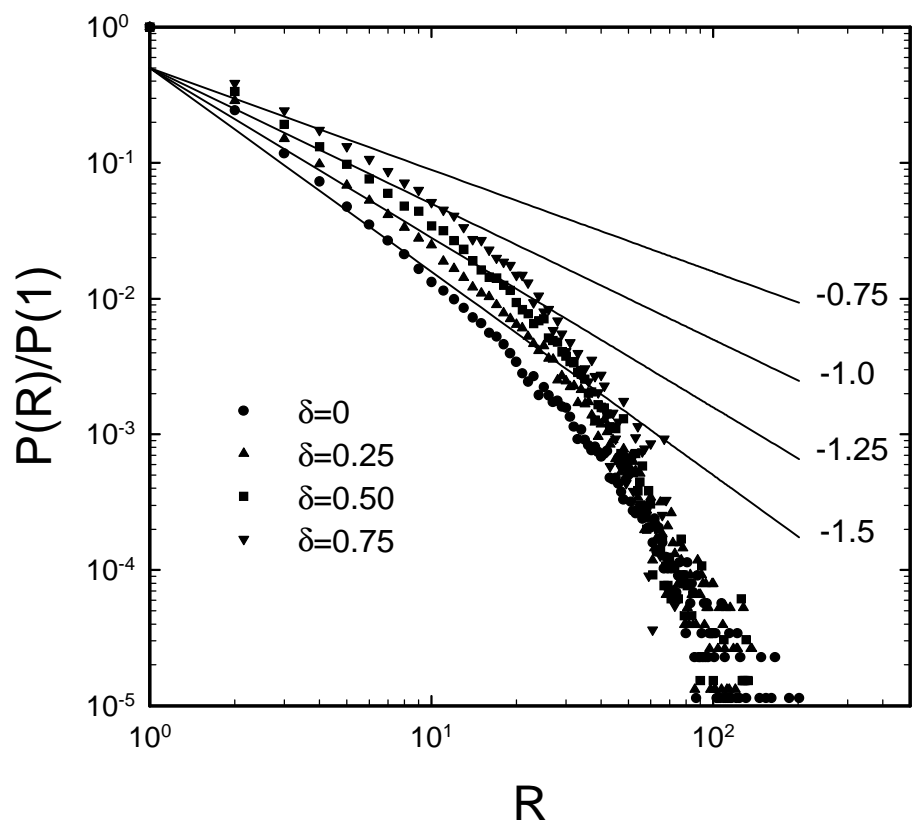
\noindent Figure 2:  The distribution of price returns $P(R)/P(1)$
as a function of $R$ on a log-log scale for different values of
$\delta$ (symbols) in Model I.
The values of $\delta$ used in the calculations are:
$\delta = 0, 0.25, 0.50, 0.75$.
The solid lines are a guide to the eye corresponding
to exponents $\alpha = 1.5, 1.25, 1.0, 0.75$ respectively.

\bigskip
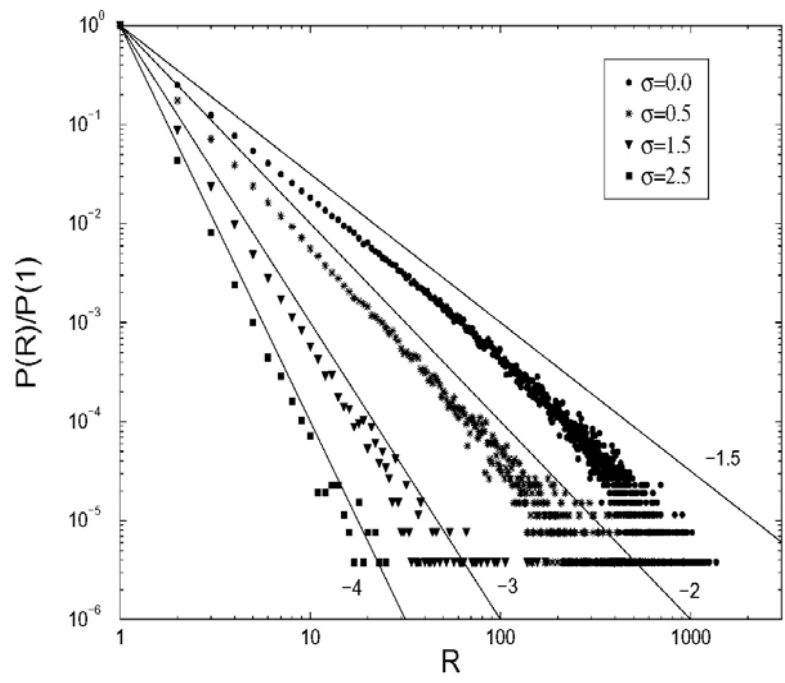
\noindent Figure 3: The distribution of price returns as a function
of $\sigma$ (symbols) in Model II.  The values of $\sigma$ used in the
calculations are: $\sigma = 0, 0.5, 1.5, 2.5$.  The solid lines are a
guide to the eye corresponding to exponents $\alpha = 1.5, 2, 3, 4$
respectively.

\end{document}